\documentstyle[12pt]{article}
\normalbaselines
\begin{document}
\bibliographystyle{unsrt}
\vbox {\vspace{6mm}}
\begin{center}
{\large \bf FIBONACCI CHAIN POLYNOMIALS: IDENTITIES FROM SELF-SIMILARITY}\\
[9mm]
Wolfdieter L a n g \\ E-mail: BE06@DKAUNI2.bitnet\\[3mm]
{\it Institut f\"ur Theoretische Physik \\ Universit\"at Karlsruhe \\
Kaiserstrasse 12, D-76128 Karlsruhe, Germany}\\[5mm]
\end{center}
\vspace{2mm}
\begin{abstract}
\noindent Fibonacci chains are special diatomic, harmonic
chains with uniform nearest neighbour interaction and two kinds of atoms
(mass-ratio $r$) arranged according to the self-similar binary Fibonacci
sequence $ABAABABA...$, which is obtained by repeated substitution of
$A\to AB$ and $B\to A$.\par
\smallskip
\noindent The implications of the self-similarity of this sequence for the
associated orthogonal polynomial systems which govern these Fibonacci
chains with fixed mass-ratio $r$ are studied.
\end{abstract}
\section{Introduction}
\hskip 1cm Fibonacci chains are linear diatomic chains with nearest neighbour
harmonic interaction of uniform strength $\kappa$ and the two masses (ratio
$r=m_{1}/m_{0}$) follow the pattern of the binary sequence
$\{h(n)\}_{1}^{\infty}$ obtained by repeated substitutions $\sigma$ in the
following way.
\begin{equation}
\sigma(1)\ =\ 10\ \ ,\ \ \sigma(0)\ =\ 1 \ \ ,
\end{equation}
starting with $0$. By definition $\sigma(uv)=\sigma(u)\sigma(v)$ for any
two strings $u$ and $v$. $\sigma^{n}(0)\equiv H_{n}$ is a string of length
$\vert H_{n}\vert = F_{n+1}$, where $F_{n}=F_{n-1}+F_{n-2}$,
$n=2,3,...$, $F_{0}=0$, $F_{1}=1$ are the Fibonacci numbers. $h(n)$ is defined
to be the $n'$th entry of the half-infinite string $H_{\infty}:= lim_{n\to
\infty}H_{n}$. {\it E.g.} $H_{5}=\sigma^{5}(0)=10110101$, $h(1)=1$, $h(2)=0$,
etc. $(1)$ is called the Fibonacci substitution rule, and the masses of the
half-infinite chain are taken to be
\begin{equation}
m_{n}=m_{h(n)}\ \ ,\ \ n=1,2,...
\end{equation}
This sequence $\{h(n)\}_{1}^{\infty}$ is self-similar because the string
$H_{\infty}$ satisfies $\sigma(H_{\infty})=H_{\infty}$. Aperiodicity follows
from this invariance, or fixed point, property. (This sequence is in fact
also quasiperiodic, but this does not concern us here.) \par\noindent
Chains of this type have been considered as models of binary alloys
\cite{A86}. For instance, one may consider chains with an elementary unit
determined by the first $N$ members of the $\{h(n)\}$ sequence and repeat it
periodically, using certain boundary conditions. This then corresponds to
$(AB)^{\infty}$ chains for $N=2$, $(ABA)^{\infty}$ chains for $N=3$, etc.\par
\noindent The dual of such chains (with equal masses but two spring constants
$\kappa_{0}$ and $\kappa_{1}$ following the pattern of the Fibonacci
substitution sequence) are related to one-dimensional quasicrystals
\cite{Sh84}. One can also make contact to artificially manufactured
superlattices \cite{M86}. \par \noindent
Originally such Fibonacci chains were considered as models for the study of
the regime in between periodic and random structures.\cite{K83,O83}.\par
\medskip \noindent
The purpose of this work is to write down the identities which are satisfied
by the characteristic polynomials of these Fibonacci chains due to the
self-similarity of the substituion sequence $\{h(n)\}$ which determines the
pattern of the masses of the oscillators. These identities will be expressed
in terms of the $2\times 2$ transfer matrices $M_{n}$ which are unimodular
and real. The matrix elements are given by the characteristic polynomials
$\{S_{n}^{(r)}(x)\}$, where $r$ is the mass-ratio of the two types of atoms
and $x$ is a normalized frequency squared
($x\equiv\omega^{2}/2\omega_{0}^{2}$, $\omega_{0}^{2}=\kappa/m_{0}$). The
zeros of $S_{N}^{(r)}(x) $ determine the eigenfrequencies of finite Fibonacci
chains with $N$ atoms and both ends fixed. One also encounters so-called
first associated polynomials $\{\hat S_{n}^{(r)}(x)\}$. They correspond to
a right shift by one unit in the substitution sequence. Hence, the zeros of
$\hat S_{N}^{(r)}(x)$ produce the eigenfrequencies of chains with masses
$m_{h(2)}=m_{0}$, ...,$m_{h(N+1)}$. Both $r-$families of polynomials
generalize Chebyshev's $\{S_{n}(y)\}$ polynomials ($S_{-1}=0$, $S_{0}=1$,
$S_{n}=yS_{n-1}-S_{n-2}$) to two variables with the identification
\begin{equation}
S_{n}^{(1)}(x)\ =\ \hat S_{n}^{(1)}(x)\ =\ S_{n}(2(1-x))\ .
\end {equation}
They constitute, for fixed mass-ratio $r$, systems of orthogonal polynomials
and have been studied in some detail in refs.\cite{L1,L2,L3,L4}.
\section{Fibonacci Chain Polynomials}
\hskip 1cm For the Fibonacci chains $(\kappa,m_{h(n)}$) defined in {\it section
the equation of motion for longitudinal, time-stationary vibrations
$q_{n}(t) =q_{n}\ exp\ (i\omega t) $ are
\begin{equation}
q_{n+1}\ +\ q_{n-1}\ -\ Y(n)q_{n}\ =\ 0\ ,\ n=1,2,...
\end{equation}
with
\begin{equation}
Y(n)\equiv 2(1-\omega^{2}/(2\omega_{n}^{2}))\ ,\ \omega_{n}^{2}\equiv \kappa/
m_{h(n)}\ .
\end{equation}
We use the two variables $r\equiv m_{1}/m_{0}$ and $x\equiv
\omega^{2}/(2\omega_{0}^{2})$. We put $Y(n)=Y$ if $h(n)=1$ and $Y(n)=y$ if
$h(n)=0$. Hence
\begin{equation}
Y\ =\ 2(1-rx)\ \ ,\ \ y\ =\ 2(1-x)\ \ .
\end{equation}
The equations of motion are rewritten with the help of the $ SL(2,\bf{R})$
transfer matrix $R_{n}$:
\begin{equation}
\left( \matrix { q_{n+1} \cr q_{n} \cr } \right ) = R_{n} \left(
\matrix{ q_{n} \cr  q_{n-1} \cr } \right) :=\left( \matrix {Y(n) &-1 \cr
 1 & 0 \cr } \right) \left( \matrix { q_{n}  \cr q_{n-1} \cr } \right)\ \
 \ .
\end {equation}
$R_{n}$ is either $R_{1}$ or $R_{0}$ depending on the $Y(n)$ value, {\it i.e}
$R_{n}=R_{h(n)}$. For the half-sided infinite chains considered here iteration
leads to
\begin{equation}
\left( \matrix { q_{n+1} \cr q_{n} \cr } \right ) = R_{n}R_{n-1}\cdots R_{1}
\left(\matrix{ q_{1} \cr  q_{0} \cr } \right) =:M_{n}\left( \matrix {q_{1} \cr
q_{0}\cr } \right) ,
\end {equation}
with the inputs $q_{1}$ and $q_{0}$ (the mass at site number $0$ is
irrelevant). $M_{n}$ is real and unimodular. The recursion
$M_{n}=R_{n}M_{n-1}$
with input $M_{1}=R_{1}$ leads to
\begin{equation}
M_{n}\ =\ \left( \matrix {S_{n} &-\hat S_{n-1} \cr S_{n-1}
&-\hat S_{n-2}\cr } \right) \ \ ,
\end{equation}
where the recursion formulae for the generalized two-variable Chebyshev
polynomials are
\begin{eqnarray}
S_{n} &=&Y(n) S_{n-1}-S_{n-2}\ \ \ \ \ \ \ \ \  ,
\ \ \ \  S_{-1}=0,
\ \  S_{0}=1\ \ \ , \\
 \hat S_{n} &=& Y(n+1) \hat S_{n-1}-\hat S_{n-2}\ \ \ \ ,\ \ \ \ \hat S_{-1}
 =0,\ \  \hat S_{0}=1\ \ \ .
\end{eqnarray}
These polynomials generate certain combinatorial numbers \cite{L5}. The
meaning of these numbers can be understood if one uses the intimate
connection of the Fibonacci substitution sequence with Wythoff's $A$
and $B$ sequences
\begin{equation}
A(n)\ =\ n\ +\ \sum_{k=1}^{n-1}h(k) \ \ \ ,\ \ \ B(n)\ =\ n\ +\ A(n)\ .
\end{equation}
These sequences $\{A(n)\}_{1}^{\infty}$ and $\{B(n)\}_{1}^{\infty}$ cover
the positive integers in a complementary way: every number $N>0$ is either
an $A-$ or a $B-$number. For an $A-$number n ({\it i.e.} $n=A(m)$ for some
$m$) $h(n)=1$, and for a $B-$number n ({\it i.e} $n=B(m)$ for some $m$) $h(n)
=0$. Wythoff's sequences are a special case of Beatty sequences:
$A(n)=\lfloor n\varphi \rfloor$,  $B(n)=\lfloor n\varphi^{2}\rfloor $,
with $\varphi^{2} =\varphi+1$, $\varphi>0$, the golden mean. \par
\noindent The characteristic polynomials $\{S_{n}^{(r)}(x)\}$, obtained from
$\{S_{n}(Y,y)\}$ by replacement of $Y$ and $y$ according to eq.(6), constitute,
for fixed mass-ratio $r$, a system of orthogonal polynomials.
$\{\hat S_{n}^{(r)}(x)\}$ are the first-associated orthogonal polynomials.
\section{Self-Similarity Identities}
\hskip 1cm The string, or 'word', $H_{\infty}$ defined in {\it section 1} is
inv
under the inverse substitution $\sigma^{-1}$, with $\sigma^{-1}(1)=0$, $\sigma
^{-1}(10)=1$. This is equivalent to the self-similarity of the sequence
$\{h(n)\}_{1}^{\infty}$ which is shown in the FIG.
\vspace{4cm}
\begin{center}
\parbox[b]{12.8cm}{
\begin{picture}(17,4)
\unitlength=1cm
\thicklines
\put (-1.3,3){\line(1,0){15.5}}
\put (-1.3,1){\line(1,0){15.5}}
\put (-1.8,3){\makebox(0,0)[b]{(l)}}
\put (-1.8,2){\makebox(0,0)[b]{$\sigma^{-1}$}}
\put (-1.8,1){\makebox(0,0)[b]{(l+1)}}
\put (14.5,3){\ldots}
\put (14.5,1){\ldots}
\put (-1,3){\line(1,-2){1}}
\put (0,3){\line(0,-1){2}}
\put (1,3){\line(0,-1){2}}
\put (2,3){\line(1,-2){1}}
\put (3,3){\line(0,-1){2}}
\put (4,3){\line(1,-2){1}}
\put (5,3){\line(0,-1){2}}
\put (6,3){\line(0,-1){2}}
\put (7,3){\line(1,-2){1}}
\put (8,3){\line(0,-1){2}}
\put (9,3){\line(0,-1){2}}
\put (10,3){\line(1,-2){1}}
\put (11,3){\line(0,-1){2}}
\put (12,3){\line(1,-2){1}}
\put (13,3){\line(0,-1){2}}
\put (14,3){\line(0,-1){2}}
\put (0,3){\circle*{.2}}
\put (3,3){\circle*{.2}}
\put (5,3){\circle*{.2}}
\put (8,3){\circle*{.2}}
\put (11,3){\circle*{.2}}
\put (13,3){\circle*{.2}}
\put (1,1){\circle*{.2}}
\put (6,1){\circle*{.2}}
\put (9,1){\circle*{.2}}
\put (14,1){\circle*{.2}}
\put (-1,3){\circle{.2}}
\put (1,3){\circle{.2}}
\put (2,3){\circle{.2}}
\put (4,3){\circle{.2}}
\put (6,3){\circle{.2}}
\put (7,3){\circle{.2}}
\put (9,3){\circle{.2}}
\put (10,3){\circle{.2}}
\put (12,3){\circle{.2}}
\put (14,3){\circle{.2}}
\put (0,1){\circle{.2}}
\put (3,1){\circle{.2}}
\put (5,1){\circle{.2}}
\put (8,1){\circle{.2}}
\put (11,1){\circle{.2}}
\put (13,1){\circle{.2}}
\put (-1,3.3){1}
\put (0,3.3){\bf 2}
\put (1,3.3){3}
\put (2,3.3){4}
\put (3,3.3){\bf 5}
\put (4,3.3){6}
\put (5,3.3){\bf 7}
\put (6,3.3){8}
\put (7,3.3){9}
\put (8,3.3){\makebox(0,0)[b]{\bf 10}}
\put (9,3.3){\makebox(0,0)[b]{11}}
\put (10,3.3){\makebox(0,0)[b]{12}}
\put (11,3.3){\makebox(0,0)[b]{\bf 13}}
\put (12,3.3){\makebox(0,0)[b]{14}}
\put (13,3.3){\makebox(0,0)[b]{\bf 15}}
\put (14,3.3){\makebox(0,0)[b]{16}}
\put (0,.2){1}
\put (1,.2){\bf 2}
\put (3,.2){3}
\put (5,.2){4}
\put (6,.2){\bf 5}
\put (8,.2){6}
\put (9,.2){\bf 7}
\put (11,.2){8}
\put (13,.2){9}
\put (14,.2){\makebox(0,0)[b]{\bf 10}}
\end{picture}
\begin{quote}
FIG.  Self-similarity of the sequence $\{h(n)\}^{\infty}_{1}$. Circles
stand for the value $1$ ($A$-numbers $n$), and disks stand for the value $0$
($B$-numbers $n$). $\sigma^{-1}(10)=1\ $, $\sigma^{-1}(1)=0\ $. Level $(l)$ is
mapped to level $(l+1)$ by $\sigma^{-1}$.
\end{quote}
}
\end{center}
The upper level, called $(l)$ in the FIG., shows the numbers marked as $A-$
and $B-$numbers. The $h(n)$ value is $1$ or $0$, denoted by a circle or disk,
respectively. When the substitution $\sigma^{-1}$ is applied one reaches the
next higher level, called $(l+1)$ in the FIG., on  which the same sequence is
reproduced. Let the position of the $n'$s number at level $(l)$ be
$x_{n}^{(l)}$ for $l=0,1,...$. Level $l=0$ is assumed to correspond to the
original sequence. Then one finds for $p\in {\bf N_{0}}\equiv{\bf N}\cup\{0\}$
\begin{eqnarray}
x_{A(p)}^{(l+1)}\ &=&\ x_{B(p)}^{(l)}\ \ , \\
x_{B(p)}^{(l+1)}\ &=&\ x_{AB(p)}^{(l)}\ \ ,
\end{eqnarray}
where $AB(p)$ stands for the composition $A(B(p))$ of Wythoff's sequences.
{\it E.g.} The number $A(4)=6$ at level $(l+1)$ occurs in the FIG. at the
same position as $B(4)=10$ at level $(l)$, or $B(2)=5$ at level $(l+1)$
corresponds to $AB(2)=A(5)=8$ at level $(l)$.\par \noindent
Iteration, depending on the parity of the level number, leads to
\begin{eqnarray}
x_{A(p)}^{(2k+1)}\ =\ x^{(0)}_{B^{k+1}(p)}\ \ &,&\ \ x_{A(p)}^{(2k)}\ =\
x_{AB^{k}(p)}^{(0)} \ \ , \\
x_{B(p)}^{(2k+1)}\ =\ x^{(0)}_{AB^{k+1}(p)}\ \ &,&\ \ x_{B(p)}^{(2k)}\ =\
x_{B^{k+1}(p)}^{(0)}\ \ .
\end{eqnarray}
Consider the level $(l+1)$ transfer matrix
\begin{equation}
M_{n}^{(l+1)}\ =\ R^{(l+1)}_{h(n)}\cdots R_{h(1)}^{(l+1)}\ \ ,
\end{equation}
satisfying the recursion relation
\begin{eqnarray}
R_{1}^{(l+1)}\ =\ R_{0}^{(l)}R_{1}^{(l)}\ \ &,& \ \  R^{(0)}_{0}\ \equiv \
R_{0}\ = \ \left( \matrix {y & -1  \cr 1 & 0 \cr} \right)\ , \\
R_{0}^{(l+1)}\ =\ R_{1}^{(l)}\ \ &,& \ \ R_{1}^{(0)}\ \equiv \ R_{1}\ =\
\left( \matrix { Y & -1  \cr 1 & 0 \cr } \right) \ \ .
\end{eqnarray}
Iteration leads , with $M_{n}\equiv M_{n}^{(0)}$, to
\begin{equation}
R^{(l+1)}_{1}\ =\ M_{F_{(l+3)}}\ \ \ \ , \ \ \ \  R_{0}^{(l+1)}\ =\
M_{F_{(l+2)}
\end{equation}
with the Fibonacci numbers $F_{n}$. \par \noindent
Due to $(15)$ and $(16)$ one has
\begin{eqnarray}
M^{(2k+1)}_{A(p)}\ =\ M_{B^{k+1}(p)}\ \ \  &,& \ \ \ M^{(2k)}_{A(p)}\ =\
 M_{AB^{k}(p)}\ \ , \\
M^{(2k+1)}_{B(p)}\ =\ M_{AB^{k+1}(p)}\ \ \  &,& \ \ \ M^{(2k)}_{B(p)}\ =
\ M_{B^{k+1}(p)}\ \ .
\end{eqnarray}
The recursion at each level is
\begin{equation}
M_{n}^{(l+1)}\ =\ R^{(l+1)}_{h(n)}\ M^{(l+1)}_{n-1}\ \ \ ,\ \ \ M_{1}^{(l+1)}\
= \ R_{1}^{(l+1)}\ .
\end{equation}
Combining iteration and recursion, in a systematic way, leads to transfer
matrix identities for level $(0)$, {\it i.e.} for the original matrices
$M_{n}$ of eq. $(8)$. One finds alltogether six families of such identities,
depending on the parity of the level one starts with and the specification
of the index. These identities are, for $m \in {\bf N}$ and $k \in {\bf N}$,
\begin{eqnarray}
(I)\qquad\phantom{xxxxx}M_{B^{k+1}(m)}\ &=& \ M_{F_{2k+1}}\ M_{AB^{k}A(m)}\ \
, \nonumber   \\
(IIa)\qquad \phantom{x}M_{B^{k}A(A(m)+1)}\ &=&\ M_{F_{2k+1}}\ M_{B^{k+1}(m)}
\ \ , \nonumber \\
(IIb)\qquad \phantom{x}M_{B^{k}A(B(m)+1)}\ &=&\ M_{F_{2k+1}}\ M_{AB^{k+1}(m)}
\ \ , \\
(III)\qquad \phantom{xxxxx}M_{AB^{k}(m)}\ &=&\ M_{F_{2k}}\ M_{B^{k}A(m)}\ \ ,
 \nonumber \\
(IVa)\qquad M_{AB^{k}A(A(m)+1)}\ &=&\ M_{F_{2(k+1)}}\  M_{AB^{k+1}(m)}\ \ ,
 \nonumber \\
(IVb)\qquad M_{AB^{k}A(B(m)+1)}\ &=&\ M_{F_{2(k+1)}}\ M_{B^{k+2}(m)} \ \ .
\nonumber
\end{eqnarray}
$(I)$, $(IIa)$, $(IIb)$ and $(III)$ result from odd levels $l=2k+1$, with
$n$ put $AB(m)$, $AA(A(m)+1)$, $AA(B(m)+1$ and $B(m)$, respectively. $(III)$,
$(IVa)$ and $(IVb)$ result from even levels $l=2k$, with $n$ put $B(m)$,
$A(A(m)+1)$ and $A(B(m)+1)$, respectively.\par\noindent
{\it E.g.} $(I)$ and $(III)$ produce for $m=1$, due to $A(1)=1$,
$B^{k+1}(1)=F_{2k+3}$ and $AB^{k}(1)=F_{2(k+1)}$, identities which are the
well-known recursion formula for transfer matrices with neighbouring
Fibonacci number indices
\begin{equation}
M_{F_{n+1}}\ =\ M_{F_{n-1}}\ M_{F_{n}}\ \ .
\end{equation}
Not all eqs.(24) are independent. {\it E.g.} if one puts $m=B(p)+1$ in $(I)$,
replaces $k$ by $k+1$ and combines it with eqs. $(IVb)$, with
$k \to k-1$ and $m\to p$, one finds eqs. $(IIa)$, due to the identiy
$B(p)+1 = A(A(p)+1)$ and eq.$(25)$ for even $n$. However, eqs.$(IIa)$ provide
identities for $M_{B^{k}A(p)}$ which complement those obtained from eqs.
$(IIb)$. \par \noindent
It is possible to combine $(I)$ of eq.$(24)$ with $(III)$ specialized to
$m\to A(m)$ and use $(I)$ again with $k\to k-1$ and $m\to A^{2}(m)$.
Continuing this process one finds for $k\in {\bf N}$ and $m\in {\bf N}$
\begin{eqnarray}
(I') \qquad  M_{B^{k+1}(m)}\ &=& \  M_{F_{2k+1}}\ M_{F_{2k}}
\cdots M_{F_{2}}\ M_{BA^{2k}(m)}  \nonumber \\
(III')\qquad M_{AB^{k}(m)}\ &=& \ M_{F_{2k}}\ M_{F_{2k-1}}\cdots M_{F_{2}}\
M_{BA^{2k-1}(m)}
\end{eqnarray}
\noindent $(I)$ and $(III)$ in $(24)$ can be replaced by both eqs. $(26)$,
and the other eqs. of $(24)$ can be rewritten using $(26)$. \par \noindent
\smallskip
The sum of the indices of the transfer matrices on the {\it r.h.s.} of
eqs.$(24)$ and $(26)$ have to match the index of the {\it l.h.s.} This
fact produces families of identities among iterated Wythoff $A$ and $B$
sequences. A detailed investigation of these Wythoff composites identities
will be given elsewhere. All of these identities can be rederived as
corollaries of a new theorem relating two seemingly different unique number
systems: the Wythoff- and the Zeckendorf- (or Fibonacci-) representations.
\par \smallskip \noindent
The transfer matrix identities $(24)$ are equivalent to those for their
matrix elements, {\it i.e.} the characteristic polynomials $\{S_{n}(Y,y)\}$
and $\{\hat S_{n}(Y,y)\}$. In order to derive them one rewrites the indices
of all matrix elements as Wythoff composites. Consider, for example, $(I)$.
For the elements of $M_{B^{k+1}(m)}$ one employs the simple identities
$B^{k+1}(m)-1=B(B^{k}(m))-1=A^{2}B^{k}(m)$ and $B^{k+1}(m)-2=ABAB^{k-1}(m)$.
The last identity can be proved for $m=A(p)$ and $m=B(p)$ separately. On the
{\it r.h.s.} of $(I)$ one rewrites the indices of the matrix elements with
the help of the identities $F_{2k+1}=B^{k}(1)$, $F_{2k+1}-1=A^{2}B^{k-1}(1)$,
$F_{2k+1}-2=ABAB^{k-2}(1)$ for $k=2,3,...$, and $F_{3}-2=0$. Moreover,
$AB^{k}A(m)-1 = BAB^{k-1}A(m)$, $\ AB^{k}A(m)-2=A^{3}B^{k-1}A(m)$. Finally,
$(I)$ decomposes into the following four sets of eqs.
\begin{eqnarray}
(I,(1,1))\qquad \phantom{xxx}S_{B^{k+1}(m)}\ &=&\ S_{B^{k}(1)}\ S_{AB^{k}A(m)}\
\hat S_{A^{2}B^{k-1}(1)}\ S_{BAB^{k-1}A(m)}\ \ , \nonumber \\
(I,(1,2))\qquad \phantom{xxx}\hat S_{A^{2}B^{k}(m)}\ &=&\ S_{B^{k}(1)}\ \hat
S_{BAB^{k-1}A(m)}\ -\ \hat S_{A^{2}B^{k-1}(1)}\ \hat S_{A^{3}B^{k-1}A(m)}\ \ ,
\nonumber \\
(I,(2,1))\qquad \phantom{xxx}S_{A^{2}B^{k}(m)}\ &=&\ S_{A^{2}B^{k-1}(1)}\
S_{AB^
-\ \hat S_{ABAB^{k-2}(1)}\ S_{BAB^{k-1}A(m)}\ \ , \\
(I,(2,2))\qquad \hat S_{ABAB^{k-1}(m)}\ &=&\ S_{A^{2}B^{k-1}(1)}\
\hat S_{BAB^{k-1}A(m)}\ -\ \hat S_{ABAB^{k-2}(1)}\ \hat S_{A^{3}B^{k-1}A(m)}\
\ . \nonumber
\end{eqnarray}
\noindent The last two sets of eqs. hold only for $k=2,3,...$. For $k=1$ one
has
\begin{eqnarray}
S_{A^{2}B(m)}\ &=&\  Y\ S_{ABA(m)}\ - \  S_{BAA(m)} \ \ , \nonumber \\
\hat S_{ABA(m)}\ &=& \ Y\ \hat S_{BAA(m)}\ - \ \hat S_{A^{4}(m)}\ \ .
\end{eqnarray}
\noindent The other eqs. in $(24)$  decompose in a similar way. The arguments
of the polynomials is always $(Y,y)$ , which can be replaced using eq.(6).
\par \noindent
This concludes the derivation of the self-similarity eqs. for the Fibonacci
chain polynomials. It is clear that further work is needed in order to extract
from this gamut of eqs. information pertaining to chain properties, like
structure of spectra and displacements.\par


\begin{thebibliography}{99}
\bibitem{A86} F. Axel, J.P. Allouche, M. Kleman, M. Mend\`es -France, and
J. Peyri\`ere, J. Physique Coll. {\bf 47}, C3 supppl. 7, 181 (1986)
\bibitem{Sh84} D. Shechtman, I. Blech, D. Gratias, and J.W. Cahn, Phys.
Rev. Lett. {\bf 53}, 1951 (1984)
\bibitem{M86} J.D. Axe, R. Clarke, R. Merlin, K.M. Mohanty, and J. Todd,
 Phys. Rev. Lett. {\bf 57}, 1157 (1986)
\bibitem{K83} M. Kohmoto, L.P. Kadanoff, and C. Tang, Phys. Rev. Lett.
 {\bf 50}, 1870 (1983)
\bibitem{O83} S. Ostlund, R. Pandit, D. Rank, H.J. Schellnhuber, and
 E.D. Siggia, Phys. Rev. Lett. {\bf 50}, 1873 (1983)
\bibitem{L1} W. Lang, J. Phys. A {\bf 25}, 5395 (1992), {\it ibid.} A{\bf
26}, 1261 (1993)
\bibitem{L2} W. Lang, in {\it Applications of Fibonacci Numbers, Vol. 5},
eds. G.E. Bergum, A. N. Philippou, and A.F. Horadam,
(Kluwer Academic Publ./Dordrecht, 1993), pp 429-440
\bibitem{L3} W. Lang, {\it Proceedings of the Third International Wigner
Symposium, Oxford, 1993}, eds. L.L. Boyle and A.S. Solomon, to be published
\bibitem{L4} W. Lang, "The Measure of the Orthogonal Polynomials Related to
Fibonacci Chains: The Periodic Case", Karlsruhe preprint KA-THEP-1-1993, 1993
\bibitem{L5} W. Lang, The Fibonacci Quarterly {\bf 30.3}, 199 (1992)
\end{thebibliography}
\end{document}